\documentstyle[times,psfig]{mn}

\title{An absorption event in the X-ray lightcurve of NGC 3227}
\author[]
  {G.~Lamer$^{1}$\thanks{E-mail: glamer@aip.de},
  P. Uttley$^2$ and I.M.~M$^{\rm c}$Hardy$^2$ \\
  $^1$ Astrophysikalisches Institut Potsdam, Germany \\
  $^2$ Department of Physics and Astronomy, The University,
  Southampton, SO17 1BJ }

\date{Accepted. Received}
\pagerange{\pageref{firstpage}--\pageref{lastpage}}
\pubyear{2001}

\pagerange{\pageref{firstpage}--\pageref{lastpage}}
\pubyear{2001}

\begin{document}

\maketitle

\label{firstpage}

\begin{abstract}
We have monitored the Seyfert galaxy NGC 3227 with the {\em Rossi
X-ray Timing Explorer}  ({\it RXTE}) since January 1999.
During late 2000 and early 2001 we observed an unusual hardening of
the 2-10 keV X-ray spectrum which lasted several months.
The spectral hardening was not accompanied by
any correlated variation in flux above 8 keV.
We therefore interpret the spectral change as transient absorption
by a gas cloud of column density $2.6 \cdot 10^{23}$~cm$^{-2}$ crossing
the line of sight to the X-ray source.  A spectrum obtained by {\em
XMM-Newton} during an early phase of the hard-spectrum event confirms the obscuration model
and shows that the absorbing cloud is only weakly ionised.  The {\it
XMM-Newton} spectrum also
shows that $\sim10$\% of the X-ray flux is not obscured, but this
unabsorbed component is not significantly variable and may be scattered
radiation from a large-scale scattering medium.  Applying the spectral
constraints on cloud ionisation parameter and assuming that the cloud
follows a Keplerian orbit, we constrain the location of
the cloud to be $R\sim10-100$~light-days from the central X-ray source,
and its density to be $n_{\rm H}\sim 10^{8}$~cm$^{-3}$, implying that
we have witnessed the eclipse of the X-ray source by a broad line
region cloud.
\end{abstract}


\section{Introduction}

The presence of broad, permitted lines in the optical spectra of
type~1 AGN, indicates the presence of dense ($n_{\rm H}>10^{8}$~cm$^{-3}$),
relatively low-ionisation gas close
to the central engine (e.g. Peterson 1997, Krolik 1999).  The covering factor of
the `broad line region' (or BLR) is estimated from line equivalent
widths to be around 10\%, while the strengths of low-ionisation and
Balmer lines imply column densities $N_{\rm
H}>10^{22}$~cm$^{-2}$ (see Peterson 1997 and Krolik 1999 for reviews).
`Reverberation mapping' measurements of time delays in the response of optical lines to
changes in the optical continuum indicate that the BLR lies within a few light-weeks of the
continuum source (e.g. Peterson et al. 1998, Kaspi et al. 2000).  

It is perhaps surprising that clearcut evidence for the low-ionisation BLR
clouds in type~1 AGN has not previously emerged in the X-ray band. 
The soft X-ray band ($<2$~keV) 
is particularly sensitive to absorption by gas along the line of sight to the
central X-ray source.   Numerous studies using missions sensitive to the soft
X-ray band (e.g. {\it ROSAT}, {\it ASCA}, {\it BeppoSAX} and most recently
the grating instruments on {\it Chandra} and {\it XMM-Newton}) have shown
the presence of absorption due to more highly ionised gas (the `warm absorber') along
the line of sight in many Seyfert 1s, but the precise location of
this gas is still fairly uncertain (e.g. M$^{\rm c}$Hardy et
al. 1995, George et al. 1998a,  Netzer et al. 2002, Schurch \& Warwick
2002).  Column density variations of cold X-ray absorbing gas
have been reported on time-scales of months-years in both type~1 and
type~2 Seyferts on time-scales of months to years 
(Malizia et al. 1997, Risaliti, Elvis \& Nicastro 2002).  If such
variations are due to clouds crossing the line of sight with Keplerian
velocities, the obscuring clouds may lie within or not much beyond the
BLR.  Unfortunately, due to the sparse temporal
sampling of X-ray spectra of AGN available with most X-ray missions, 
it is difficult to put these apparent absorption changes into context, and
assess if they really are caused by the motion of dense absorbing
clouds across the line of sight to the X-ray source.

In this letter, we present a study of the long-term X-ray 
spectral variability of the Seyfert 1.5 galaxy NGC~3227, using data from a 
monitoring campaign carried out by the {\it Rossi X-ray Timing Explorer} 
({\it RXTE}).  A previous study of spectral variability in NGC~3227
showed evidence of an order of magnitude increase
in the cold/low-ionisation column (from $\sim10^{21}$~cm$^{-2}$ to 
$\sim10^{22}$~cm$^{-2}$) between {\it ASCA} observations obtained
in 1993 and 1995 (George et al. 1998b).  Using data from
our monitoring campaign, we show that in 2000/2001 NGC~3227 underwent an
unusual event lasting $\sim3$~months, during which the spectrum
became exceptionally hard.  The symmetry of this event suggests it is
due to absorption by a high column density cloud 
($N_{\rm H}\sim3\times10^{23}$~cm$^{-2}$)
moving across the line of sight to the X-ray source.  This 
interpretation is confirmed by an {\it XMM-Newton} observation obtained
serendipitously, early in the event, which shows that the cloud is not
substantially ionised.  We use dynamical arguments,
along with constraints on the ionisation state of the cloud, to infer
that the cloud most likely originates in the BLR.

\section{Observations and data reduction}
\label{obs}

We have been monitoring NGC 3227 with {\em RXTE} since January 1999.
The observations were carried out as pointings of duration about 1 ksec.
The frequency of the monitoring varied between weekly observations
to four observations per day.  
Here we present data from the Proportional Counter Array (PCA),
which consists of 5 Xenon filled proportional counter units
(PCUs) and is sensitive to X-rays in the 2-60 keV band.
For each pointing we extracted the Xenon layer 1  
data from all PCUs that were switched on during the observation, and
extracted spectra (and model background spectra) according to the
standard method outlined by Lamer et al. (2002).
PCA response matrices were calculated invididually for each
observation using {\sc pcarsp v2.37}, taking into account temporal
variation of the detector gain and the changing numbers of detectors
used.

In order to increase the signal to noise ratio of the PCA spectra,
we added all spectra within weekly intervals.
Each of these weekly spectra was fitted in {\sc xspec 11.0.1} with a model
consisting of a power law, a Gaussian emission line, and the Wisconsin
absorption model. The energy range used for spectral 
fitting was restricted to 2-12 keV. 
Since the data were extracted from varying combinations of
PCU's and the detector gain was changed on several occasions
during the time span of the monitoring campaign, simple count rate
lightcurves cannot be used to measure the variability of the source.
We therefore calculated model fluxes from the spectral fits 
in the bands 2-10 keV and 8-10 keV in order to obtain 
detector independent light curves over the full time range
of the monitoring campaign.

NGC 3227 was observed by {\em XMM-Newton} on
28-29 November 2000. We retrieved the EPIC and RGS pipeline
products from the XMM science data archive (XSA).
Using the XMM Science Analysis System (SAS v5.3)
we produced  EPIC MOS and PN spectra for NGC 3227.
The extraction radius was 30'' and only events in  
single and double patterns with $FLAG=0$ were selected.

\section{A transient absorption event}
\label{spectral}
\subsection{{\it RXTE} monitoring}

\begin{figure}
 \par\centerline{\psfig{figure=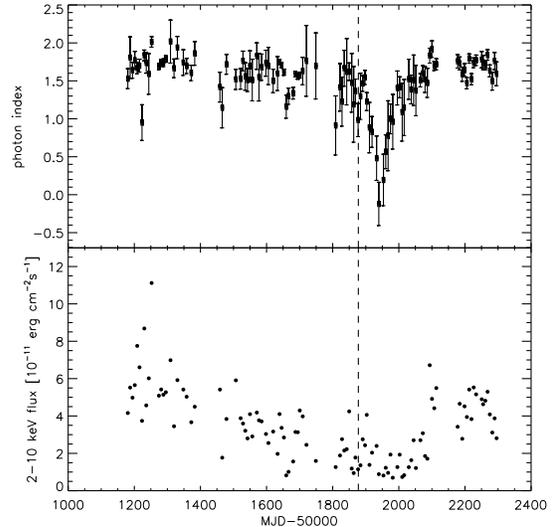,width=7.5truecm}}
 \caption{\label{indexcurve}
   Top: Variability of the spectral index for power law models 
        with the absorbing column density fixed to the galactic value.
   Bottom: 2-10 keV flux lightcurve. The dashed line indicates the epoch
   of the {\em XMM} observation. 
   }
\end{figure} 

\begin{figure}
 \par\centerline{\psfig{figure=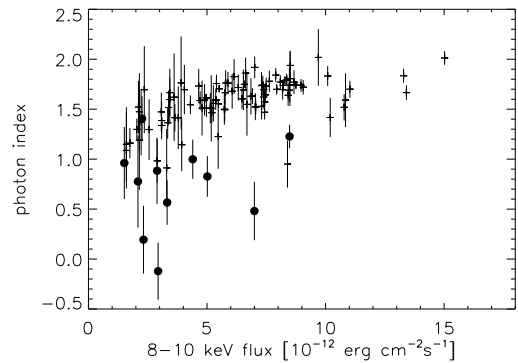,width=7.5truecm}}
 \caption{\label{fluxgamma}
    Power law spectral index vs. 8-10 keV flux during the 
    epoch of the hard-spectrum event MJD=51900-52000 (full dots) and
during all other times (crosses).
   }
\end{figure}

During the first quarter of the year 2001, transient hardening 
of the 2-12 keV spectrum was observed. The event lasted for
about 100 days and peaked around MJD 51950.
Due to the limited spectral resolution and signal to noise 
ratio of the spectra the spectral fits alone cannot distinguish
whether the hardening is caused by intrinsic slope variation
or by additional absorption. 
When we fit the weekly binned spectra with the power law + Gaussian
line model and the absorbing column density set to the galactic 
value of  $N_{\rm H}=2.2\cdot 10^{20} {\rm cm^{-2}}$, we find 
a flattening of the photon index to $\Gamma \sim 0$,
harder than so far observed in any AGN X-ray spectrum (see Fig. \ref{indexcurve}).

We now consider the two simplest possibilities: that the transient
hardening is caused by a change in the primary continuum shape, or
that it is caused by obscuration by a cloud of large column density. 
Since the hard-spectrum event occurs during a period when the source
long-term X-ray flux is quite low, and the X-ray spectra of a
number of Seyfert~1 are known to become harder at low fluxes
(e.g. Uttley et al. 1999, Vaughan \& Edelson 2001, Lamer et al 2003), we first test
whether the spectral shape is simply correlated with flux.  If this
hypothesis is true, we would expect the relation between photon index
and flux to be the same both within the 100~d hard-spectrum event and at other times.
Specifically, we need to measure photon index versus 8-10~keV flux
(i.e. where absorption effects are minimal), in order to remove any
spurious correlation produced by varying absorption,
which would naturally cause the flux in softer bands (which are
absorbed) to correlate with the fitted photon index.
Fig. \ref{fluxgamma} shows the photon indices of the weekly binned spectra plotted
against the 8-10 keV flux.  We find that the 'normal' spectral variability of NGC~3227
is very similar to that found in other Seyfert galaxies.
At low flux levels the spectral index steepens rapidly with 
increasing flux; at higher flux levels the increase of the spectral
slope levels off and saturates at  $\Gamma \sim 2.0$.
Similar flux - slope relations have been found in NGC 4051
(Lamer et al. 2002) and  MCG-6-30-15 (Shih, Iwasawa \& Fabian 2002).
On the other hand, the data points from the 2001 event clearly
fall outside the normal range in the flux - index plane.
The indices are much harder and do not strongly correlate with changes
in the 8-10 keV flux.

\begin{figure}
 \par\centerline{\psfig{figure=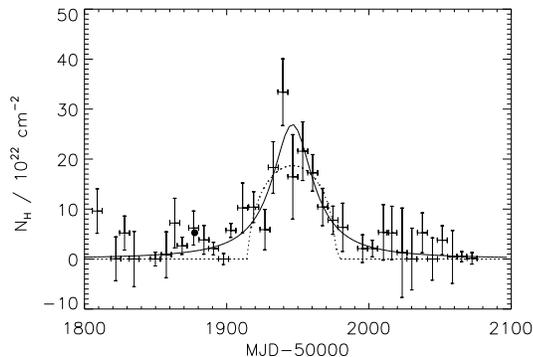,width=7.5truecm}}
 \caption{\label{nhcurve}
   Best fit absorbing column densities $N_{\rm H}$ from power law fits
   the {\em RXTE} spectra. The   photon index was fixed at $\Gamma=1.6$.
   The dotted and solid lines correspond to best-fitting uniform sphere and $\beta$ models respectively. 
The full dot marks the $N_{\rm H}$ measurement derived from the
{\em XMM-Newton} EPIC spectra. 
   }
\end{figure}

Having ruled out simple, flux-correlated variability as the source of
the hard-spectrum event, we next consider the possibility that it is caused by
transient obscuration by a large column of gas.  The apparent symmetry
of the event supports this possibility, suggesting that a symmetric
(i.e. spherical) cloud passed in front of the X-ray source.
From the flux of hard X-rays and 
the normal flux - index relation in NGC 3227 (Fig. \ref{fluxgamma}) 
we estimate an intrinsic photon index of $\Gamma = 1.6$ during the
absorption event.
We repeated the model fitting with the power law + Gaussian line
model with the photon index fixed at  $\Gamma = 1.6$ and
the absorption column density left free to vary.
Fig.~\ref{nhcurve} shows the resulting column density for
the assumption of a neutral absorber.  We fitted the profile of column
density changes with two simple models, a uniform sphere and a
$\beta$-model of the form $N_{\rm H}=N_{\rm H,max}/(1+(r/R)^2)^{0.5}$ (see Fig.~\ref{nhcurve}).  
The  $\beta$-model gave a much better fit (reduced
$\chi^{2}\simeq0.98$ versus $\chi^{2}\simeq1.85$ for the uniform sphere
model), for a $\beta$ index of $\sim0.5$, suggesting that the cloud
density is not uniform but increases towards the centre. 
The column density peaks at   $N_{\rm H} \sim 2.6 \cdot 10^{23}$~cm$^{-2}$.
However, based on the {\it RXTE} data alone, we cannot absolutely discount the
possibility that the hard-spectrum event is caused by very unusual
variability of the primary continuum, which is not simply correlated
with flux.
Fortunately, NGC~3227 was observed by {\it XMM-Newton}
early in the event, and it is to this data that we now turn.

\subsection{The {\it XMM-Newton} spectrum}

\begin{figure}
 \par\centerline{\psfig{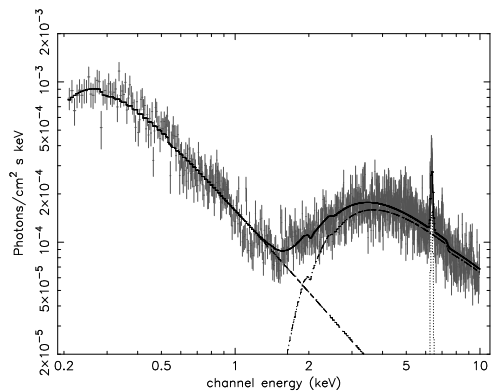}}
 \caption{\label{xmmspec}
Unfolded best-fitting {\em XMM-Newton} EPIC PN spectrum fitted with a
double power-law plus warm absorber and Gaussian (model~3, see text for details).
   }
\end{figure}

NGC 3227 was observed by  {\em XMM-Newton} around MJD 51877,
during an early phase of the hard spectrum event (see Fig \ref{nhcurve}). 
We performed joint model fitting on the EPIC MOS1, MOS2 and PN spectra.
The hard bands ($>$ 2 keV) can be modeled by a strongly absorbed  
($N_{\rm H} \sim 5 \cdot 10 ^{22} {\rm cm^{-2}}$)
power law spectrum and a narrow Fe ${\rm K_{\alpha}}$ fluorescence line.
The absorbing column density is consistent with the value measured 
by {\em RXTE} at the same time (see Fig. \ref{nhcurve}).
In the soft range below 1 keV an unabsorbed power law spectrum dominates
(see below and Fig.~\ref{xmmspec}).

Using {\sc xspec}, we fitted 4 different partial covering models to the {\em XMM} EPIC spectra:

1. Single power law model with partial absorption:
  {\sc wabs(powl + zwabs(powl + zgauss))}, with the two power law components having identical slopes.

2. Double  power law model:
   {\sc wabs(powl1 + zwabs(powl2 + zgauss))}, both power law indices are free to vary.

3. Double power law model with warm absorber as partial coverer:
   {\sc wabs(powl1 + absori(powl2 + zgauss))}

4. Single power law with Compton reflection and warm absorber:
 {\sc wabs(powl + absori(hrefl(powl) + zgauss))}

\begin{figure}
 \par\centerline{\psfig{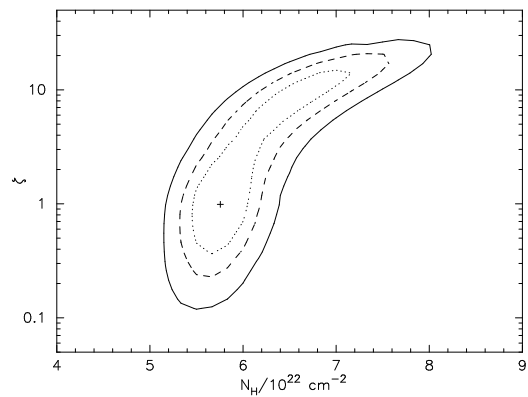}}
 \caption{\label{xinhcont}
 Constraints on ionisation parameter $\xi$ and absorbing column
$N_{\rm H}$ from fits to the {\em XMM-Newton} spectrum using model 3
(see text).  Dotted, dashed and solid lines correspond to 68\%, 90\%
and 99\% confidence contours respectively.  The best-fitting values
are marked by a cross.
   }
\end{figure}

The results of the spectral fitting are summarized in table
\ref{specfit}, and the double power law + warm absorber model is shown in Figure~\ref{xmmspec}.
In all of the three models a narrow iron ${\rm K_{\alpha}}$ fluorescence 
line with rest frame energy 6.4 keV and equivalent width $\sim$ 240 eV
is required.  The neutral, unredshifted absorption in all three models is
consistent with the Galactic value ($2.2\times10^{20}$~cm$^{-2}$).
From  model 3, we find that constraints on $N_{\rm H}$ and the 
ionisation parameter, $\xi$ of the absorbing gas suggest that the
obscuring cloud is not neutral, but ionised, albeit with a low
ionisation parameter (see Fig.~\ref{xinhcont}). 

Note that the absorbed power-law component is rather hard
($\Gamma\sim1.3$), as expected due to the low 8-10~keV flux
($2.2\cdot10^{-12}$~erg~cm$^{-2}$~s$^{-1}$) measured
during the {\it XMM-Newton observation} (see Fig.~\ref{fluxgamma} for
comparison with the `normal' photon index-flux relation).
The unabsorbed power law component in the soft range 
has a normalization of about 10\% of the absorbed power law component.
The best fit parameters of models 2 and 3 (also see
Fig.~\ref{xmmspec}) also show that the unabsorbed component is intrinsically 
softer than the absorbed component. There are three possible reasons for this finding:
1. We see two components emitted from different regions of the AGN, e.g. the softer,
unabsorbed power law arises from a more extended emitting region,
which is not completely obscured by the cloud while the inner, hard
X-ray emitting region is obscured.
2. The absorbed and unabsorbed components are intrinsically identical, 
but have a steeper slope in the soft range below $\sim 1$~keV. This could be due 
to a soft excess or a hard reflection spectrum at higher energies. The reflection spectrum (model 4)
fits the data well with a rather high ratio ($3.2\pm0.6$) of reflection covered to escaped component.
The large  reflected fraction may be due to the low flux state during the time of the XMM observations.  
3. The slope of the underlying power law spectrum is highly variable, and therefore
the slope of the scattered and delayed component differs from the slope of the 
aborbed, nuclear component.

\begin{figure}
 \par\centerline{\psfig{figure=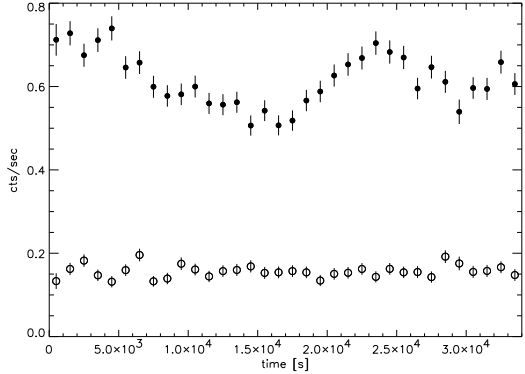,width=7.5truecm}}
 \caption{\label{lightcurve}
  {\em XMM} EPIC PN lightcurve in the 0.2-1 keV band (open circles) and in the
      2-10 keV band (full dots).        
   }
\end{figure}

We produced {\em XMM} EPIC  light curves for both the unabsorbed component in the 0.2-1 keV range
and for the absorbed component at 2-10 keV (Fig. \ref{lightcurve}). While
the absorbed emission is clearly variable, no significant variability is found in the soft
component (a constant model yields $\chi^{2}=36.7$ for 33 d.o.f.). 
We therefore conclude that the two components are emitted in different
regions of the AGN. Since the soft power-law component does not show any intra-day variability,
it could arise from a relatively large volume, and might represent
radiation scattered from an extended, ionised medium.

We note that, since submission of this paper the Nov 2000 XMM-Newton
observation of NGC 3227 has also been published by Gondoin et al. (2003).
They fit a slightly different model to us, we refer readers to Gondoin et al. for a 
full discussion of their model.

\begin{table*}
\centering
\caption{\label{specfit} {\em XMM} EPIC spectral fits} 
\begin{tabular}{lcccccccc}
 & & \multicolumn{2}{c}{\em Unabsorbed Component} &
\multicolumn{4}{c}{\em Absorbed Component} & \\ 
model & $N_{\rm H,1}$ & $\Gamma_1$ &  norm$_1$ & $N_{\rm H,2}$ & $\Gamma_2$ & norm$_2$ & $\xi$ & $\chi^2 (dof)$ \\
      &                &           &  $10^{-4}$  &               &          & $10^{-4}$ &        &           \\
      &  $10^{20}{\rm cm^{-2}}$ & &  ${\rm cm^{-2}\;s^{-1}\;keV^{-1}}$ & 
         $10^{20}{\rm cm^{-2}}$ & &  ${\rm cm^{-2}\;s^{-1}\;keV^{-1}}$ & 
         ${\rm erg\;cm\;s^{-1}}$ & \\ 
1 powl
      &$0.0 ^{+0.2}_{-0.0}$&$1.45\pm0.02$&$1.62\pm0.03$&$632\pm9$&$1.45\pm0.02$&$17.4\pm1.0$& - & 1707 (1565) \\
2 powl 
      &$1.4\pm0.4$    &  $1.71\pm0.06$ & $1.64\pm0.05$&$527\pm16$&$1.28\pm0.03$&$13.0\pm1.3$&- & 1673 (1564) \\
warm abs.
      &$1.6\pm0.5$  &  $1.74\pm0.06$ &
$1.64\pm0.06$&$576\pm33$&$1.29\pm0.03$&$13.4\pm1.4$&$1.0^{+1.5}_{-0.6}$& 1650 (1562)\\
warm abs. + refl.&$1.5\pm0.4$&$1.72\pm0.05$&$1.64\pm0.05$&$708\pm35$&$1.72\pm0.05$&$22.5\pm2.2$&$0.5_{-0.3}^{+0.8}$& 1647 ( 1562) \\    
\end{tabular}
\end{table*}

\section{Physical constraints on the absorbing matter} 
\label{absorber}

Assuming that the hard-spectrum event is caused by the passage of an absorbing
cloud along the line of sight to the X-ray source, we can place
lower limits on the distance of the cloud from the source and its
density by making a few basic assumptions.

For simplicity, we shall assume a cloud of uniform density and 
ionisation parameter, following a circular orbit around the X-ray
source.   Although our assumption that the cloud has a
uniform density is almost certainly not true (see
Figure~\ref{nhcurve}), the discrepancy in $N_{\rm H}$
profiles between the $\beta$-model
and uniform sphere model is typically only a factor of a few.
Therefore we do not expect any significant (i.e. order of
magnitude) error in our estimates of cloud parameters to result from
this simplifying assumption.  The ionisation parameter $\xi$ is given by

\begin{equation}
\xi=\frac{L_{\rm ion}}{n_{\rm H}\;R^{2}},
\end{equation}

where $L_{\rm ion}$ is the ionising luminosity in the range
13.6~eV--13.6~keV, $n_{\rm H}$ is the hydrogen number density of the gas
and $R$ is its distance from the ionising source.  The diameter of the
cloud is given by $N_{\rm H}/n_{\rm H}$, so that the
velocity of the cloud across the line of sight is
$v_{\rm cloud}=N_{\rm H}/(n_{\rm H}t_{\rm cross})$ (where $t_{\rm cross}$
is the crossing time of the cloud across the line of sight).  Assuming a
circular orbit of radius $R$ around the central black hole of 
mass $M_{\rm BH}$, we can obtain an expression for the hydrogen number
density:

\begin{equation}
n_{\rm H}=\frac{N_{\rm H}}{t_{\rm cross}}\;\sqrt{\frac{R}{G\;M_{\rm BH}}}.
\end{equation}

where $G$ is the gravitational constant.
Finally, we can combine equations 1 and 2 and rearrange 
to obtain $R$:

\begin{equation}
R\simeq 4\cdot10^{16} M_7^{\frac{1}{5}}\left( \frac{L_{42}\;t_{\rm
days}}{N_{22}\;\xi} \right) ^{\frac{2}{5}}\;\;\;\;\;{\rm cm},
\end{equation}

where $L_{42}=L_{\rm ion}/10^{42}$~erg~s$^{-1}$, $t_{\rm days}$ is 
the crossing time in days, $M_{7}=M_{\rm BH}/10^{7}$~M$_{\odot}$ and
$N_{22}=N_{\rm H}/10^{22}$~cm$^{-2}$.  We next take the conservative 99\%
confidence limits imposed on ionisation parameter by the {\it
XMM-Newton} spectrum of $\xi =$0.1-30 (see Fig.~\ref{xinhcont}),
a cloud maximum column density of
$N_{\rm H}=1.87\cdot10^{23}~cm^{-2}$ and crossing time of 62~days (the best fitting
uniform sphere parameters).
The best estimate of black hole mass from reverberation mapping
is $M_{\rm BH}=5\cdot10^{7}$~M $_{\odot}$ (Wandel, Peterson \& Malkan 1999), 
and we estimate $L_{\rm ion}=10^{42}$~erg~s$^{-1}$
($H_{0}=70$~km~s$^{-1}$~Mpc$^{-1}$, using the observed flux during the event and 
assuming a typical continuum photon index $\Gamma=1.6$).  Combining
these values in equation 3, we
calculate that $R$ lies in the range 9-86~light-days. Using these limits in equation~2, we find a
corresponding range of possible densities, $n_{\rm H}\sim0.7-2.0\times10^{8}$~cm$^{-3}$, so
that the cloud must be of order a light-day in diameter or less.  Note
that these estimates are not strongly dependent on the central black
hole mass, to which no strong lower limit can be set in NGC~3227
(Wandel et al. 1999).  A factor 10 reduction in black hole mass will only
reduce the lower and upper limits on $R$ by 37\% and increase
$n_{\rm H}$ by 250\%.

\section{Discussion}
\label{discuss}
We now consider how the physical constraints we can place on the
properties of the eclipsing cloud in NGC~3227 compare with the
constraints expected for gas in various regions of the AGN.  Most obviously,
the location of the cloud (roughly 10-100 light-days from the central
X-ray source) is consistent with an origin in the BLR of NGC~3227, the size of which is estimated 
from optical reverberation studies to be between 10-20~light-days (Wandel et al. 1999,
Winge et al. 1995, Salamanca et al. 1994), although note that since these estimates refer
to the region of the BLR which emits most of the Balmer lines, the
full extent of the BLR may be significantly larger.
Similarly, the density we estimate for the cloud is also consistent
with the lower limit on densities of BLR clouds estimated from optical
line ratios (e.g. see Krolik 1999 for discussion) and rules out an
origin in the lower density, more distant, narrow line
region (NLR) or large-scale molecular torus envisioned by the standard unification
model.  This last constraint is particularly strong, since even in
the absence of any lower bound on ionisation parameter, the density of
the cloud can only increase with assumed orbital radius, due to the well constrained
total column density and the strong dynamical constraint on cloud size imposed by the
cloud crossing time.  

By the same token, the cloud cannot be very dense, as according to our
equation 2, the density
scales only with the square root of distance from the central X-ray
source, so if the assumed density were significantly larger than $10^{8}$~cm$^{-3}$
the cloud would be located much further from the X-ray source than is
consistent with any known component of the AGN (the same reasoning can
be used to strongly constrain cloud size).  Our strong constraint on
cloud density is in conflict with the estimate of much denser BLR clouds ($n_{\rm
H}\sim10^{11}$~cm$^{-3}$) in the luminous Seyfert~1 NGC~5548, by
Ferland et al. (1992), who required a large hydrogen number density to
prevent the clouds becoming too heavily ionised in the high radiation
density environment implied by the relatively small BLR size estimated
from reverberation mapping.  However, we note that although NGC~3227 has
a similar sized BLR to NGC~5548 (e.g. see Peterson et al. 2002), it has a much lower X-ray luminosity
($10^{42}$~erg~s$^{-1}$ versus  $5\times10^{43}$~erg~s$^{-1}$), so that the
radiation density in the BLR of NGC~3227 is likely to be much lower than in
NGC~5548.  In fact, if for comparison, we convert from our ionisation
parameter $\xi$ to the equivalent formulation $U$ used by Ferland et al. (1992) (e.g. see George et
al. 1998a for conversion factors), we find $U<0.1$
in NGC~3227, consistent with the value estimated for
NGC~5548 and the BLR in other AGN (see Ferland et al. 1992, Peterson
1997, and references therein).

The size inferred for the eclipsing cloud is quite large, implying
that only of order $\sim100$ clouds are required to produce the $\sim10$\%
covering fraction estimated for the BLR (Peterson 1997, Krolik 1999).
This small number of clouds conflicts with the large number
($>10^{4}$) required to produce the smooth broad line profiles observed in type 1~AGN
(e.g. Atwood, Baldwin \& Carswell 1982), assuming that the cloud intrinsic
line widths are small, thermal widths only.   This
latter assumption might be relaxed if the cloud line widths are
intrinsically broadened, e.g. if the clouds are dynamically unstable,
which might be expected given the large cloud size.
Note also that our cloud size estimates can be used to constrain the
size of the primary X-ray emitting region to be less than
$\sim1$~light-day, or around 170~Schwartzschild radii for a $5\cdot
10^{7}$~M$_{\odot}$ black hole (assuming
the primary X-ray emitting region is totally covered, as
suggested by the lack of variability in the unabsorbed spectral
component).

When considering the implications of these results for models of the
BLR in AGN in general, one must bear
in mind the caveat that strong observational selection effects almost
certainly exist in any detection of BLR clouds by X-ray absorption.
For example, eclipses of the X-ray emitting region by much smaller
clouds, of the higher densities envisaged by Ferland et al. (1992),
would only be noticable if the X-ray emitting region was itself very
compact (smaller than hundreds of light-seconds or less than one Schwartzschild
radius for a $5\cdot10^{7}$~M$_{\odot}$ black hole).  Furthermore, the
duration of such events would be much shorter than the observed
$\sim100$~d (perhaps only a few thousand seconds), so would only be
detectable with continuous monitoring, which we do not have.  Therefore we cannot rule out the
existence of such small clouds in NGC~3227.  Other selection effects
may also be at work.   For example, the relatively low radiation
density in the BLR of NGC~3227 may permit the existence of large,
relatively low-density clouds which might not exist in other AGN.
The presence of such clouds may also account for the smaller amplitude column-density variations
reported in NGC~3227 by George et al. (1998b).  Alternatively, the fact that
column variations are apparently common in NGC~3227 might be explained if the BLR has
a planar geometry and we are viewing the X-ray source through the
BLR plane.  Even taking into account the above caveats, our results imply that a
population of large BLR clouds exists in at least some AGN.  Taking
the estimate of much smaller, higher density clouds in NGC~5548
(Ferland et al. 1992) at face value,  it seems that BLR cloud densities and sizes in different
AGN must span at least a three order-of-magnitude range.

\section*{Acknowledgments}

We would like to thank Hagai Netzer and Brad Peterson for
helpful discussions.

\label{lastpage}

\end{document}